\def\kms{\ifmmode{\,\hbox{km}\,s^{-1}}\else {\rm\,km\,s$^{-1}$}\fi}

\def\msun{{\rm\,M_\odot}}
\def\lsun{{\rm\,L_\odot}}

\def\kmsm{{\rm\,km\,s^{-1}\,Mpc^{-1}}}

\def\hmpc{\ifmmode{h^{-1}\,\hbox{Mpc}}\else{$h^{-1}$\thinspace Mpc}\fi}

\def\eg{{\it e.g.}~}

\def\et{{\it et~al.}~}

\def\sigp{\ifmmode{\sigma_p}\else {$\sigma_p$}\fi}
\def\sig1{\ifmmode{\sigma_1}\else {$\sigma_1$}\fi}
\def\r200{\ifmmode{r_{200}}\else {$r_{200}$}\fi}
\def\m200{\ifmmode{M_{200}}\else {$M_{200}$}\fi}

\documentstyle[11pt,aaspp4]{article}
\tighten

\slugcomment{DRAFT: \today}

\begin{document}

\title
{$\Omega_b$ via Oort's Method}

\author
{
R.~G.~Carlberg\altaffilmark{1,2},
S.~L.~Morris\altaffilmark{1,3},
H.~K.~C.~Yee\altaffilmark{1,2},
E.~Ellingson\altaffilmark{1,4},
R.~Abraham\altaffilmark{1,3,5},
H.~Lin\altaffilmark{1,2},
D.~Schade\altaffilmark{1,3},
P.~Gravel\altaffilmark{1,2},
C.~J.~Pritchet\altaffilmark{1,6},
T.~Smecker-Hane\altaffilmark{1,3,7}
F.~D.~A.~Hartwick\altaffilmark{6}
J.~E.~Hesser\altaffilmark{3},
J.~B.~Hutchings\altaffilmark{3},
\& J.~B.~Oke\altaffilmark{3}
}

\altaffiltext{1}{Visiting Astronomer, Canada--France--Hawaii Telescope, 
        which is operated by the National Research Council of Canada,
        le Centre National de Recherche Scientifique, and the University of
        Hawaii.}
\altaffiltext{2}{Department of Astronomy, University of Toronto, 
        Toronto ON, M5S~3H8 Canada}
\altaffiltext{3}{
        National Research Council of Canada,
        Herzberg Institute of Astrophysics,     
	Dominion Astrophysical Observatory, 
        5071 West Saanich Road,
        Victoria, BC, V8X~4M6, Canada}
\altaffiltext{4}{Center for Astrophysics \& Space Astronomy,
        University of Colorado, CO 80309, USA}
\altaffiltext{5}{Institute of Astronomy, 
        Madingley Road, Cambridge CB3~0HA, UK}
\altaffiltext{6}{Department of Physics \& Astronomy,
        University of Victoria,
        Victoria, BC, V8W~3P6, Canada}
\altaffiltext{7}{Department of Physics \& Astronomy,
        University of California, Irvine,
        CA 92717, USA}


\begin{abstract}
The baryon density of the universe is equal to the product of the
baryon-to-light ratio, $M_b/L$, and the luminosity density, $j$. We
estimate $M_b/L$ as the sum of the masses of the X-ray gas and the
visible stars in a rich cluster of galaxies divided by the luminosity
of the cluster galaxies in precisely the same sky aperture. We
evaluate the gas-to-light ratio derived from the EMSS detect cell flux
and the CNOC cluster redshift survey galaxies. After making an
aperture correction to an effective overdensity of $500\rho_c$, we
find that $\Omega_{gas}=0.012-0.016 h^{-3/2}$, depending on the galaxy
fading correction. Adding in the galaxy baryons at a mass-to-light
ratio of $5\msun/\lsun$, equivalent to $\Omega_\ast=0.003h^{-1}$, we
find that $\Omega_b=0.015-0.019$ for $H_0=100 \kmsm$ (or $0.040-0.051$
for $H_0=50$). Expressed as the baryon to photon ratio, $\eta$, this
corresponds to $\eta=4.0-5.2\times 10^{-10}$ ($H_0=100$) and is in the
mid-range of values from other methods.  The individual clusters have
a dispersion about the mean $\Omega_{gas}$ of 40\%, and the $\chi^2$
of the 14 clusters is consistent with the hypothesis that the
gas-to-light ratio is a universal constant.  If we ignore the light of
the cD, the variance increases by a factor of three.  After the radial
segregation of gas and light within a cluster is taken into account,
these statistics indicate that there is little variation of the
gas-to-light ratio from cluster to cluster over the 0.2 to 0.55 range
in redshift.
\end{abstract}


\clearpage
\section{Introduction}

The baryon density of the Universe, $\Omega_b=\rho_b/\rho_c$, is a
fundamental cosmological parameter. The ratio of the baryon density to
the photon density in the cosmic background radiation,
$\eta=273\times10^{-10} \Omega_bh^2$ (for the observed CBR
temperature), effectively measures the matter-anti-matter asymmetry at
a very early time, likely when the universe was (re-)heated in the
earliest moments of the Big Bang. Much later in the expansion of the
universe, when it is a few minutes old, nuclear reactions assemble the
light elements. The theory of Big Bang Nucleosynthesis (BBN) predicts
the abundances of the light elements given the single parameter,
$\eta$, and a knowledge of the number of types of
neutrinos. Consequently, given observations that can be used to infer
the primordial abundances of H, its isotope D, He and Li, the BBN
theory predicts a value of $\Omega_b$ for consistency.  The current
measurements imply $\Omega_b=0.007-0.024h^{-2}$
(\cite{walker,copi,hogan,st}).  The Helium abundances point toward
relatively low values of $\Omega_b$ (\eg\
\cite{steigman,hos}). Deuterium, as seen in conjunction with Hydrogen
in high redshift absorption line systems, is currently a somewhat
controversial indicator. Depending on the details of line
identification and strength measurements, Deuterium favours either an
$\Omega_b$ similar to that obtained from Helium
(\cite{songaila,carswell}) or a value near the upper end of the
confidence range (\cite{tfb}). For the values of $\Omega_b$ under
discussion, the primordial Lithium abundance favours the low to mid
$\Omega_b$ range but falls near a minimum of the predicted abundances
so has reduced power to discriminate.

Another approach to $\Omega_b$ measurement via Lyman $\alpha$
clouds can be used to estimate $\Omega_b$. These measurements rely on
further assumptions, mainly the ionizing flux and the velocity
distribution of the absorbing gas, which are constrained through other
data and models. The result of the two model analyses is that
$\Omega_bh^2\ge 0.017$, which is consistent with the high $\Omega_b$
(\cite{rauch,weinberg}).

High velocity dispersion galaxy clusters have long been recognized as
powerful indicators of cosmological parameters.  Measurements of the
galaxy distribution and velocity dispersions in clusters show that the
galaxies and the total mass have statistically identical distributions
over the virialized region (\cite{profile,br,ave}). Furthermore, there
is relatively little differential evolution between cluster and field
galaxies (Lin \et\ in preparation, \cite{profile}).  In these
controlled circumstances one can use with confidence Oort's method
(\cite{oort,gt}) to estimate the mass density parameter of the
universe, $\Omega_M = M/L \times j/\rho_c$, where $M/L$ is the total
cluster mass to light ratio, corrected for differential evolution, $j$
is the field luminosity density and $\rho_c$ the critical density. The same
method can be used to measure $\Omega_b=(M_{gas}+M_{stars})/L \times
j/\rho_c$. In both cases one measures only those components of the
mass that are sufficiently cool that they cannot escape the
gravitational field of 1000 \kms\ clusters.

The ratio of the gas mass to the total mass inside some radius in a
cluster, $f_g$, (hereafter referred to as the gas-to-mass ratio),
calculated or normalized over the full virialized volume of the
cluster, is expected to be nearly equal to the cosmic mean value,
although this remains to be empirically established.  It has long been
known that $f_g$ is 10-30\% (\eg\ \cite{ku,es}), but only recently has
the precision of both $f_g$ measurements and the BBN prediction become
sufficient that a clear discrepancy between $f_g$ and
$\Omega_b/\Omega_M$ with $\Omega_M=1$ emerged (\cite{wnef,wf,myers}).
The approach to measuring $\Omega_b$ here is similar to that used by
Steigman \& Felten (1995, see also
\cite{myers}), but has the innovation that we bypass the problem of
measuring the cluster mass. That is, we use Oort's method, $\Omega_b =
M_b/L \times j/\rho_c$.  This is a direct measurement of $\Omega_b$
with minimal assumptions and a minimal error budget.  Because
$\Omega_M$ is also estimated with Oort's method, our $\Omega_b$ and
$\Omega_M$ will move up or down together in the presence of a common,
undetected, systematic error.

A critical consistency test of the assumption that $\Omega_b$ is
faithfully represented by cluster baryons is that the $\Omega_b$ of
individual clusters does not vary beyond the statistical confidence
range. The situation with the gas-to-mass ratio is not clear at the
moment, although it seems likely that much of the variance comes from
the various complications in estimating masses (\cite{wf}). Here we
can examine the variance of the ratio of cluster gas mass to cluster
luminosity (hereafter the gas-to-light ratio) within our observed
region to test for consistency with a universal ratio.

The next section of the paper describes our calculation of the gas
mass within the EMSS ``detect cell''. Because the X-ray luminosity
depends on the square of the emitting mass, it is relatively
insensitive to the details of the emission modeling.  The total light
within the same aperture, both with and without the cD included, is a
straightforward sum over the observed galaxies. The resulting
gas-to-light ratio needs to be corrected for a well known aperture
effect.  In Section 3 we calculate the product of the gas-to-light
ratio with the luminosity density at the same mean redshift to give
the gas component of the baryons, to which we add our estimate of the
stellar baryons to give the $\Omega_b$ of all the visible baryons. We
conclude with a discussion of the strengths and weaknesses of this
result and the prospects for improvement. All calculations in this
paper take $H_0=100 h \kms$ and $q_0=0.1$.

\section{EMSS/CNOC Cluster Gas-to-Light Ratios}

The Canadian Network for Observational Cosmology (CNOC) redshift
survey cluster sample (\cite{yec}) was drawn from those Einstein
Medium Sensitivity Survey (EMSS) clusters with $L_x\ge 1\times
10^{44}$ erg~s$^{-1}$ in the 0.3 to 3.5 Kev passband. We restricted
the CNOC sample to the redshift range $0.18\le z \le 0.55$ for
observational reasons.  

The CNOC survey measured velocities of the cluster galaxies and their
Gunn $r$-band luminosities (\cite{global}). The EMSS observational
quantity is the X-ray flux in the ``detect cell'', a 2.4 arcminute
square aperture which is centered on the peak of the X-ray flux.  In
this section we describe how we use the X-ray and optical data to
measure a gas-to-light ratio.

\subsection{Cluster Centers and Temperatures}

The cluster centers have both X-ray and optical estimators, which
should be consistent with the same location for the calculation of the
gas-to-light ratio.  The X-ray centers of Gioia \et\ (1990) are
defined as the peak of the X-ray flux and have an accuracy of 50
arcseconds. We adopt the location of the Brightest Cluster Galaxy
(BCG) as the center of the optical light (\cite{global,profile}) at
the co-ordinates given in Gioia \& Luppino \et\ (1994) which have an
accuracy of about 5 arcseconds.  the X-ray and BCG co-ordinates have a
mean difference of 33\arcsec.  The largest deviation comes from
MS0906+11 which has a low significance difference of 58\arcsec.  We
conclude that the X-ray centers and the BCG centers are statistically
consistent with each other for our sample of rich clusters with large
X-ray luminosities.  We were unable to obtain a reliable velocity
dispersion for the cluster MS0906+11 (\cite{global}), so we will not
include it in any of the following analysis.

The galaxy velocities are in equilibrium with the cluster potential
and their distribution is consistent with tracing the mass
distribution (\cite{profile,br}). If the X-ray gas is also in
equilibrium and traces the potential, then the galaxies should have an
equivalent temperature equal to the RMS X-ray temperature, $T_x$,
derived from the spectrum. The velocity dispersion $\sigma_1$ implies
a temperature $\Re T_\sigma/\mu = \sigma_1^2$, where $\Re$ is the gas
constant and $\mu$ is the mean molecular weight of the gas. We find
$T_\sigma = 10^8 (\sigma_1/1180 \kms)^2$~K for a gas that is 74\%
Hydrogen and 25\% Helium by mass.  The X-ray temperatures of nine of
the CNOC clusters have been derived by Mushotzky and Scharf (1997,
hereafter MS97). For the sample as a whole they find that the quantity
$\beta =T_\sigma/T_x$ has a mean statistically equal to unity, as has
been seen in other samples (\eg\ \cite{lubin}).

To refine the agreement between $T_x$ and $T_\sigma$ we test whether
the agreement is consistent within the errors, that is, besides
predicting the mean of the distribution, we check for excess variance
above the errors in the individual $\beta$ values. We compare the ASCA
derived $T_x$ from MS97 (converting 1 Kev to $1.16\times10^7~^\circ$K)
to the $T_\sigma$ estimated from the CNOC velocity dispersions.  The
error in $\beta$, $\epsilon_\beta$, is calculated from quadrature sum
of the jacknife errors of the velocity dispersions (\cite{global}) and
half of the MS97 confidence range of X-ray temperatures.  The cluster
with the greatest deviation, $2.1\epsilon_\beta$, is MS1455+22, which
also stands out in the richness-velocity dispersion relation (Yee,
private communication). In both cases the indication is that the
cluster's true velocity dispersion is closer to $\sim900$
\kms\ than to the $1170\pm150 \kms$ that we found
(\cite{global}). There are no strong indications that anything is
particularly amiss, but there is no question that more cluster
velocities would help clarify the situation for this cluster. For the
nine clusters the average $\beta=0.95\pm0.10$. The distribution has
$\chi^2=15.3$ (or 11.1 without MS1455+22) which is about 9\% (20\%)
probable, hence at this level we cannot reject the hypothesis that all
$\beta$ are equal to unity. Without MS1455+22 the mean
$\beta=0.87\pm0.08$. This is possibly a very weak indication that the
X-ray gas is somewhat hotter than the virial temperature of the total
mass distribution. If true, this might be consistent with the gas
being somewhat more extended than the mass (\cite{wnef,djf}).

\subsection{Converting the Aperture Flux to a Gas Mass}

The EMSS survey measures the X-ray flux, $f_x$ in the 0.3-3.5Kev band
in a 2.4 arcminute square on the sky. To determine the gas mass that
is emitting this radiation we proceed as follows.  The emitted
luminosity in the 0.3$(1+z)$ to 3.5$(1+z)$ Kev band is $L_x = 4\pi
d_L^2(z,q_0) f_x$, where $d_L(z,q_0)$ is the luminosity distance to
the cluster.  Above, we established that $T_\sigma$ accurately
predicts the X-ray temperature, so we will use $T_\sigma$ to calculate
the volume X-ray emissivity, $n_en_H\varepsilon(T_\sigma)$, of the 14
EMSS/CNOC clusters.  We will assume that the gas visible within the
detect cell is isothermal.  The spectral emissivity is calculated
using the publicly available\footnote{\raggedright Version with last
modification Sept 21, 1993, from
ftp://heasarc.gsfc.nasa.gov/software/plasma\_codes/raymond}
Raymond-Smith code (\cite{rs}). We sum over the 0.3$(1+z)$ to
3.5$(1+z)$ Kev bins to give $\varepsilon(T_\sigma)$.  The cluster
metal abundances are taken to be 0.4 of the Allen (1973) values, to match
approximately the metal abundances inferred for clusters.  The mass
density is $\rho=1.24\times 10^{-24}\sqrt{n_en_H/0.6}$ gm~cm$^{-3}$.

We must assume a density profile for the cluster gas. A form which
accurately fits most cluster's inferred X-ray gas profile is
(\cite{jf}),
\begin{equation}
\sqrt{n_en_H}(r) = {{n_0 }\over{1+(r/a)^2}},
\label{eq:denr}
\end{equation}
where $a$ is the core radius of the gas distribution, measured to be
about $0.125\hmpc$ (\cite{jf}). We use $a= 0.125 (\sigma_1/1000\kms)
\hmpc$ which takes into account the nearly linear increase of all
scales with velocity dispersion (\cite{nfw}).  The exact value of the
core radius makes relatively little difference to the derived gas
mass, typically about a 50\% change for a factor of 4 in $a$, which is
much greater than the expected range of $a$ variation.

The X-ray luminosity from the part of the cluster in the detect cell
is the projection along the line of sight of
$\varepsilon(T)n_en_H(r)$, integrated over the detect cell, which is a
square with sides of physical length $2b$. That is, $L_x =
8\int_0^b\int_0^b\int_0^\infty n^2(r)
\,dx\,dy\,dz, $ where $r=\sqrt{x^2+y^2+z^2}$. With the aid of the
Maple symbolic integrator,
\begin{equation}
L_x = n_0^2\varepsilon(T) \pi a^3 
\Big[
\arctan{\Big({{ab}\over{b\sqrt{2b^2+a^2}+b^2+a^2}}\Big)} 
+\arctan{\Big({{a}\over{\sqrt{2b^2+a^2}}},
	{{b\sqrt{2b^2+a^2}-b^2-a^2}\over{b\sqrt{2b^2+a^2}}}}\Big)\Big],
\label{eq:lx}
\end{equation}
where $\arctan{(y,x)}$ has a range of $[-\pi,\pi]$ to give the result in
the correct quadrant.  Given the measured $L_x$, we use
Eq.~\ref{eq:lx} to derive $n_0$, the central gas density of the
cluster.  Then the gas mass projected in the detect cell is
\begin{equation}
M_{gas} = 4\pi  \rho_0a^2 b \int_0^1 \hbox{\rm arcsinh}{\Big(
{1\over{\sqrt{x^2+(a/b)^2}}}\Big)}\,dx,
\end{equation}
which is easily integrated numerically. 

The inferred central gas density, $n_0$, of Equation~\ref{eq:denr}, is
shown against the redshift of the cluster in Figure~\ref{fig:den}.
The values are consistent with what is seen in other samples at low
redshift. Not surprisingly, there is no detectable evolution
(\cite{sigma8,mush}) nor clear correlations with other quantities.

\subsection{The Gas-to-Light Ratio}

It is straightforward to sum the selection function weighted
luminosities of the galaxies within the bounds of the detect cell over
the redshift range which contains the cluster galaxies.  We have
previously demonstrated that the galaxy number profile, as selected in
the k-corrected Gunn $r$ band (hence reasonably insensitive to the
color differences between cluster and field galaxies) has a
distribution which is statistically identical to the total mass
distribution (\cite{profile,br}).  There is detectable evolution of
both the cluster and field galaxies over the redshift range we have
observed (\cite{schade_e,schade_d}, Lin \et\ in preparation).  The
Gunn $r$ data are acceptably described for ($q_0=0.1$) with pure
luminosity evolution and no density evolution, at the rate of about
$1\pm0.5$ magnitude per unit redshift. In this system the field
luminosity density, integrated to $L=0$ using the fitted luminosity
function (as is the cluster, so the correction has no net effect), is
constant and is equivalent to $\rho_c/j = 1543\pm283 h\msun/\lsun$.

The Schechter luminosity function fits find that cluster and field
galaxies are brightening at approximately at the same rate with
redshift. However the galaxy populations are different and we must
allow for the differential evolution between cluster and field
galaxies. Most of the cluster galaxies were former field galaxies that
were accreted onto the cluster with effectively no starbursting, but a
rapid suppression of the star formation (\cite{a2390,balogh}).  By
computing average luminosities we find that cluster galaxies are faded
about $0.11\pm0.05$ mag in Gunn $r$ relative to the field
(\cite{global}) which we adopt as our correction. A slightly larger
fading of about $0.3\pm0.1$ mag is seen in the rest frame B band (Lin
\et\ in preparation) as is to be expected
for the 0.2 mag average color difference between field and cluster
galaxies (for details, see \cite{profile}).  However, because the
bright galaxy population within these very rich clusters is similar to
these galaxies, the ``detect cell'' galaxies have properties similar
to the cluster galaxies as a whole.  Another approach to estimating
the differential evolution is to model the evolution of the galaxies
in color and magnitude. Using the Bruzual \& Charlot GISSEL package
(1993 and revisions) we find that for the observed $\Delta(g-r)_z=0.2$
mag color difference between the field and the cluster the expected
fading is $\Delta M_r\simeq 0.4$ mag for galaxies with a fairly wide
range of star formation histories prior to their termination in the
cluster.  The difference between the fading from the model and the
measured luminosity difference could be taken as an indication that
cluster galaxies are about 30\% more massive than field galaxies,
possibly as a result of merging, however this requires further
examination.  Except for this differential evolution correction our
survey measures the luminosities of field and cluster galaxies at the
same time so that the selection functions and most corrections are in
common and simply cancel in the $\Omega$ calculation.

The fractional errors of the gas-to-light ratios are similar numbers,
whereas the absolute errors correlate with the values themselves.  We
must therefore calculate the averages and variances using the
logarithms of the gas-to-light ratios. The $\chi^2$ of the
$\Omega_{gas}$ values about their mean is $\chi^2=20.8$, which for 13
degrees of freedom is about 8\% probable, suggesting that there may be
a further source of variance beyond the error distribution alone.
There are no significant correlations in these data between
$\Omega_{gas}$ and $L_x$ or $\sigma_1$, however, there is a
significant correlation with redshift.  The natural physical
interpretation of this redshift correlation is that it is an
``aperture effect''. Clusters at low redshift are known to have an
increasing gas-to-mass ratio with increasing radius
(\cite{djf,wf}). There is strong evidence that clusters at fixed $T_x$
or $T_\sigma$ have little evolution in their X-ray properties with
redshift (\cite{sigma8,mush}).  Moreover, the CNOC clusters have a
constant mass-to-light ratio (evolution corrected luminosities,
\cite{global,ring2}). Hence, we would expect that clusters at higher
redshift, which have a larger physical radius within the fixed angular
size of the detect cell aperture, will have higher $\Omega_{gas}$
values than those at low redshift.

The aperture correction for gas-to-light ratio is relatively small and
stable. We do not try to extrapolate either the gas mass or the light
to larger radii, but only seek to normalize our results to a mean
interior overdensity, $\delta$, of $500\rho_c$ (\cite{evrard}) using a
simple fit to the gas-to-total mass profiles of David, Jones \& Forman
(1994).  From their derived cluster gas to mass profiles we estimate
that $f_g(500) = f_g(\delta)/[1-0.35\log{(\delta/500)}]$.  The
calculated overdensities range from about 700 to about 7000.  The
resulting average correction is about 30\%, with the largest
correction being 66\%. With this correction the $\chi^2=16.8$ for 13
degrees of freedom, which is entirely consistent with no variance
beyond the errors. This correction is an important element which will
be better determined from X-ray imaging studies of these clusters
(Lewis \et\ in preparation).

In Figure~\ref{fig:om} we display the corrected $\Omega_b =
\Omega_{gas}+\Omega_\ast$. We have estimated the stellar baryons
as having $\Omega_\ast=0.003h^{-1}$, on the basis of an average
mass-to-light ratio of approximately $5 \msun /\lsun$ (\cite{mb}) and
our closure value of approximately $\simeq 1500 h \msun /\lsun$.  The
average $\Omega_b=0.019\pm0.02$ for our measured 0.11 mag luminosity
differential between cluster and field.  This $\Omega_b$ is equivalent
to an $\eta=4.0\times 10^{-10}$ (for $h=1$). Although it is a
``mid-range'' value, it is within the statistical errors of Tytler,
Fan and Burles (1996).  For the 0.4 mag of fading, our results are
just beyond their $2\sigma$ confidence range, and comparable to the
lower $\Omega_b$ generally derived from Helium (\cite{hos,st}).

\section{Discussion and Conclusions}

The average of the 14 $\Omega_b$ values is $0.019$ if the fading of
field to cluster galaxies is the 0.11 mag, as derived from the
luminosities of the galaxies in our sample, or, $\Omega_b=0.015$,
based on the predicted 0.4 mag fading derived from stellar population
modeling of the color difference between the field and cluster
galaxies. These values are in the mid range of the current values, but
remains consistent with the high values (\cite{tfb}).  The random error
is 12\%, considerably less than the systematic uncertainties.  For
$H_0=50\kmsm$ $\Omega_b=0.040-0.051$.  The individual clusters are
corrected for the average internal segregation of gas and mass in
which the gas is more extended than the mass.  Our $\Omega_b$ value is
calculated for clusters at a mean redshift of 0.31, taking
$\Omega_M=0.2$ and $\Omega_\Lambda=0$.  If this low density universe
has an $\Omega_\Lambda=0.8$, the $\Omega_b$ are reduced about
24\%. The random error in the result is 12\% whereas the potential
unresolved systematic errors are about 30\% and dominate the error
budget.  The distribution of the corrected gas-to-light ratio values
about the mean has $\chi^2=16.8$ for 13 degrees of freedom, which is
consistent with there being no intrinsic variation of the gas-to-light
ratio from cluster to cluster, beyond the mean variance of 40\%.  This
reinforces one of the benefits of the gas-to-light ratio estimator,
which is that projection effects have relatively little effect. The
lack of variation in the aperture corrected gas-to-light ratio from
cluster to cluster, over quite a large range in redshift, constrains
any significant possibilities for a large scale segregation of gas and
light (or by extension, gas and mass) external to the clusters.  The
greatest weakness of this measurement is that the X-ray fluxes are not
measured much beyond the cores of the clusters. X-ray data extending
to larger radii (Lewis \et\ in preparation) will reduce the aperture
correction and reduce the error in estimating the enclosed optical
luminosity.

Eventually an overall consistency of the cosmological parameters as
measured from various sources will pin them down, and the biases in
their measurement. For instance, if the relatively low value of
$\Omega_b\simeq0.012$ from Helium (\cite{hos}) is accepted as the
correct value, then we would conclude that correction for the fading
of field galaxies to cluster galaxies is the 0.4 mag estimated from
the colors and consistency with the luminosity functions likely
requires that cluster galaxies are on the average about 30\% more
massive (from, say, merging) than field galaxies. A test of this will
soon be possible using the CNOC field sample of galaxy groups whose
galaxies much more closely resemble the general field population.  The
errors in the $\Omega_{gas}$ values will soon be reduced as X-ray
imaging data become available for these clusters.

To test whether the baryons in the stars of the cD are uncorrelated
with the gas in the central region, we omit the light of the cD galaxy
from the gas-to-light ratio. The variance of the $\Omega_b$ values
then increases by a factor of three, such that the $\chi^2$ strongly
indicates that the gas-to-light ratios are showing a significant
variation.  We conclude that the cD is statistically coupled to the
central gas of the cluster. Moreover, given the stability of the
gas-to-light ratio from cluster to cluster and with redshift, the
conversion efficiency of gas to stars (in cluster galaxies) is
constant from place to place, within the 40\% errors of our
measurement.

Many of the elements of the calculation of $\Omega_b$ are in common
with our calculation of $\Omega_M$, which found
$\Omega_M=0.19\pm0.06$. The alternate field-to-cluster fading of 0.4
mag would lower this value to $\Omega_M=0.15\pm0.05$.  Any as yet
uncovered systematic error in common to the two will cause the two
values to rise or fall together. Consequently, the possibility that
$\Omega_M=1$ seems remote, since it would demand that either
$\Omega_b$ is about 5 times larger than we find here or that there be
a very substantial segregation between dark matter and hot gas,
opposite in sign to what would be expected. That is, the dark matter,
generally supposed to be cold on the basis of the power spectrum of
density fluctuations (\eg\
\cite{pd}) be kept out of these high velocity dispersion clusters,
whereas the X-ray plasma, known to be at least as hot as the cluster
mass field, would have to be retained or even enhanced within the
clusters.

\acknowledgments

We thank Bill Forman and Richard Mushotzky for advice on X-ray plasmas
We thank CFHT for the technical support which made the optical
observations feasible.  NSERC and the NRC provided financial support.

\clearpage

\figcaption[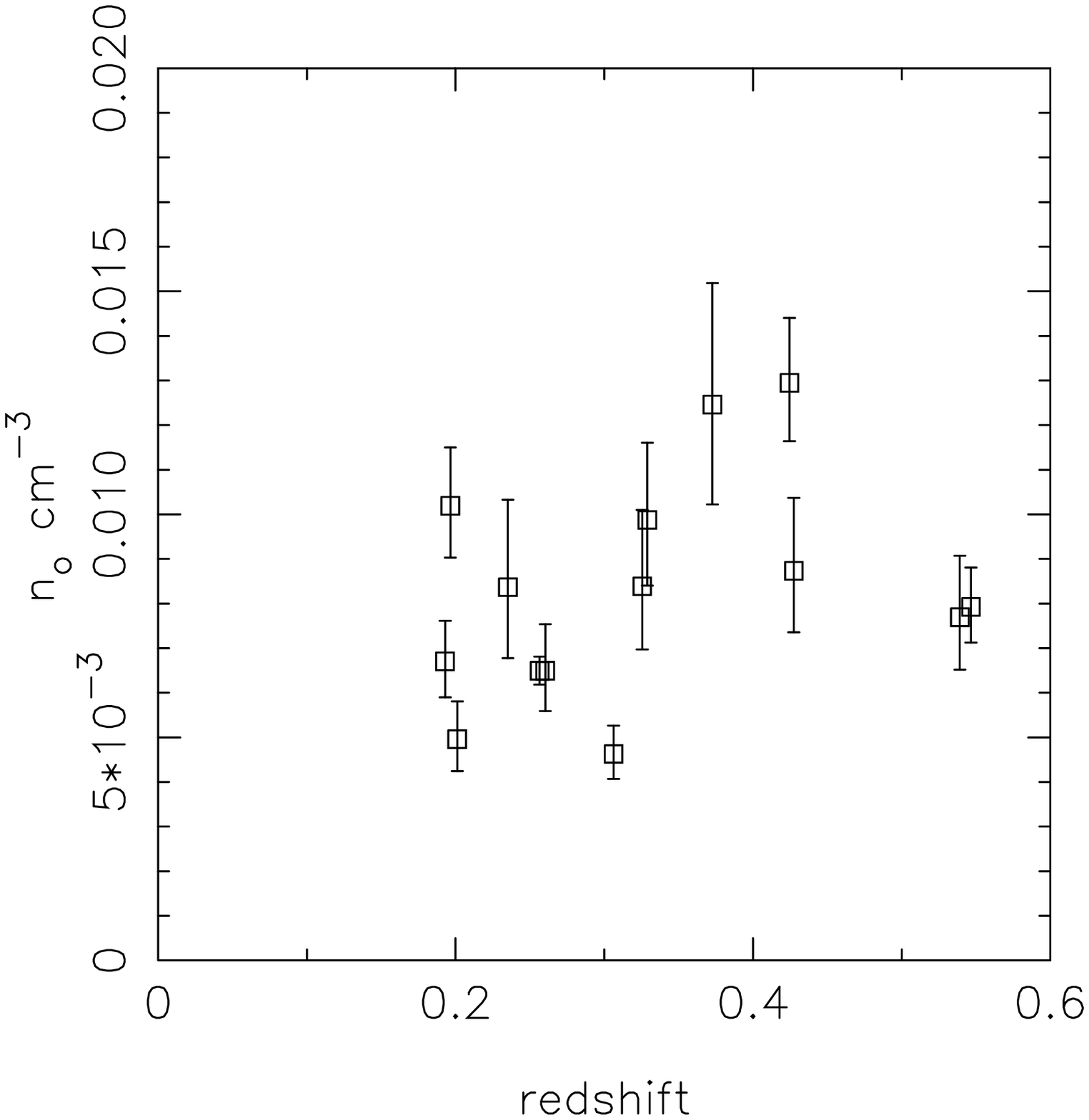]{
The central density, $n_0$, inferred from the X-ray flux for the 14
EMSS/CNOC clusters versus the redshift.
\label{fig:den}}

\figcaption[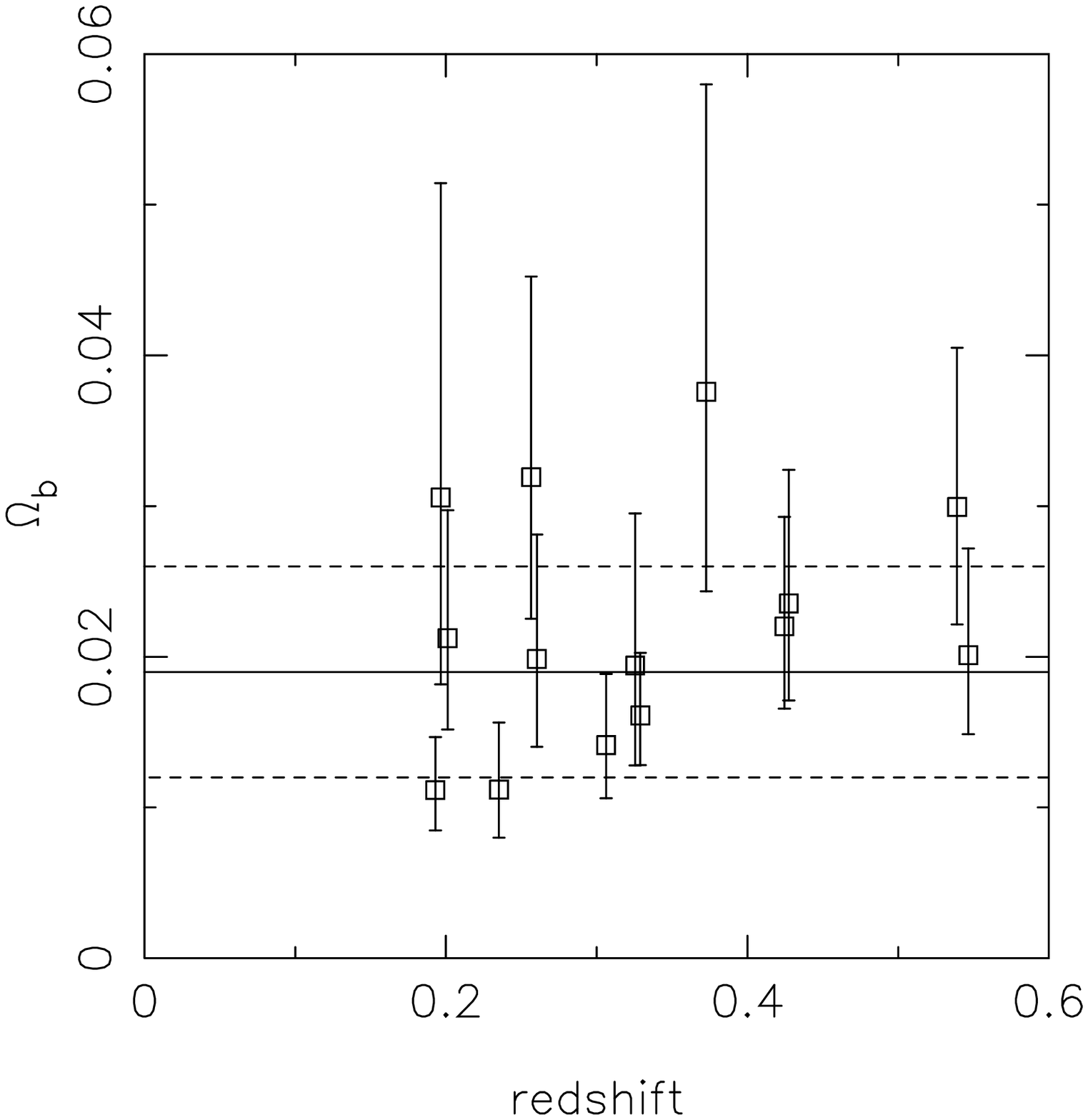]{The aperture corrected estimate of 
$\Omega_b$ with the stellar mass included, plotted against the cluster
redshift. This distribution is consistent with a single universal
value of $\Omega_b$. The variance weighted mean of the logarithmic
distribution, $\Omega_b=0.019$ and the population variance are shown
for a field to cluster fading of 0.11 mag.  The error in the mean is
about 12\%.
\label{fig:om}}

\begin{figure}[h] \figurenum{1}\plotone{fig1.ps} \caption{}\end{figure}  
\begin{figure}[h] \figurenum{2}\plotone{fig2.ps} \caption{}\end{figure}  

\end{document}